\documentclass[aps,prb,twocolumn,groupedaddress,showpacs,superscriptaddress,amssymb,amsmath]{revtex4-1}
\usepackage{graphicx}
\usepackage{amsmath}
\usepackage{amssymb}
\usepackage{amsfonts}
\usepackage{color}
\usepackage{bm}        
\usepackage{comment}
\usepackage{placeins}
\usepackage[mathscr]{eucal}
\usepackage[colorlinks, linkcolor=red,citecolor=blue,urlcolor=blue]{hyperref}
\usepackage[normalem]{ulem}
\usepackage[all]{hypcap}
\usepackage{multirow}
\usepackage[dvipsnames,usenames,table]{xcolor}
\usepackage{dcolumn}
\usepackage{hyperref}
\usepackage{bm}
\usepackage{epsf}
\usepackage{subfigure}
\usepackage{amsfonts}
\usepackage{epstopdf}%
\newcommand{\be}{\begin{equation}}
\newcommand{\ee}{\end{equation}}
\newcommand{\bea}{\begin{eqnarray}}
\newcommand{\eea}{\end{eqnarray}}
\newcommand{\ket}[1]{\vert #1 \rangle}

\begin{document}
\title{Experimental study of the thermodynamic uncertainty relation}  
\author{Soham Pal}
\affiliation{Department of Physics,
		Indian Institute of Science Education and Research, Pune 411008, India}
\author{Sushant Saryal}
\affiliation{Department of Physics,
		Indian Institute of Science Education and Research, Pune 411008, India}
		\author{D. Segal}
\affiliation{Department of Chemistry, University of Toronto, Toronto, Ontario, Canada M5S 3H6}
\affiliation{Department of Physics, University of Toronto, Toronto, Ontario, Canada M5S 1A7}
\author{T. S. Mahesh}
\affiliation{Department of Physics,
		Indian Institute of Science Education and Research, Pune 411008, India}
\author {Bijay Kumar Agarwalla}
			\email{bijay@iiserpune.ac.in }
\affiliation{Department of Physics,
		Indian Institute of Science Education and Research, Pune 411008, India}
		
\date{\today}

\begin{abstract}
A cost-precision trade-off relationship, the so-called thermodynamic uncertainty relation (TUR),
has been recently discovered in stochastic thermodynamics.
It bounds certain thermodynamic observables in terms of  the associated entropy production.  
In this work, we experimentally study the TUR in a two-qubit system using an NMR setup. 
Each qubit is prepared in an equilibrium state, but at different temperatures.
The qubits are then coupled, allowing energy exchange (in the form of heat). 
Using the quantum state tomography technique
we obtain the moments of heat exchange within a certain time interval and analyze the relative uncertainty 
of the energy exchange process.
We find that generalized versions of the TUR, which are based on the fluctuation relation, are obeyed. 
However, the specialized TUR, a tighter bound that is valid under specific dynamics,
is violated in certain regimes of operation, in excellent agreement with analytic results.
Altogether, this experiment-theory study provides a deep understanding of heat exchange in quantum systems,
revealing favorable noise-dissipation regimes of operation.
\end{abstract}

\maketitle 


{\it Introduction.}
Obtaining universal bounds of experimentally accessible physical observables has been a fundamental topic in physics. 
Such bounds include the Heisenberg uncertainty relation of quantum mechanics, 
Carnot bound for the efficiency of heat engines and Landauer erasure principle 
stemming from the second law of thermodynamics.  
Likewise, recent studies have shown that for systems that are out-of-equilibrium, 
there exist trade-off relations between the relative uncertainty of integrated currents (heat, charge)  
and the associated entropy 
production \cite{Barato:2015:UncRel,Gingrich:2016:TUP,Polettini:2016:TUP,Pietzonka:2016:Bound,Hyeon:2017:TUR,Horowitz:2017:TUR,Proesmans:2017:TUR,Gingrich:2017,Pietzonka:2017:FiniteTUR,Garrahan:2017:TUR,Dechant:2018:TUR, Pietzonka:2017:FiniteTUR,Falasco,SamuelssonM,Garrahan:2017:TUR,Dechant:2018:TUR,Koyuk:2018:PeriodicTUR,Garrahan18,Sasa:TUR,Udo:TURB,SamuelssonQP,Goold,Gabri,Vu,Hasegawa1,Hasegawa2,Saito,TUR-gupta, TURQ,BijayTUR,JunjieTUR,SamuelssonQP, Horowitz:2019:TUR,Pigolotti:TURF,Landi-PRL,Hwang,Mayank,Bio,Interacting,Van,Hyst,TUR-bijay1,Passage,VanTUR}.
These results are now collectively refereed to as \textit{Thermodynamic uncertainty relations} (TUR). 
The specialized version of the TUR (S-TUR) reads, 
\be
\frac{\langle Q^2 \rangle_c}{\langle Q \rangle^2} \geq \frac{2}{\langle \Sigma \rangle},
\label{eq:fund-TUR}
\ee
where $Q$ represents any integrated current, such as heat or charge, and it is a stochastic variable. 
$\langle Q \rangle, \langle Q^2 \rangle_c$ are the average integrated current and its noise, respectively, 
and $\langle \Sigma \rangle $ is the net average entropy production in the heat exchange process, 
characterizing irreversibility, or how far the system is driven away from equilibrium.
The S-TUR was first conjectured for continuous time, discrete state Markov process 
in steady state \cite{Barato:2015:UncRel}. It was later proved with the large deviation technique 
\cite{Gingrich:2016:TUP,Horowitz:2017:TUR}. Since then, this relation has been generalized to discrete time, 
discrete state Markov process \cite{Proesmans:2017:TUR}, 
finite time statistics \cite{Dechant:2018:TUR,Pietzonka:2017:FiniteTUR,Horowitz:2017:TUR,Pigolotti:TURF}, 
Langevin dynamics \cite{Dechant:2018:TUR,Dechant:2018:TUR,Gingrich:2017,Hasegawa1,TUR-gupta,Hyeon:2017:TUR},
 periodically driven systems \cite{Koyuk:2018:PeriodicTUR,Gabri}, 
multidimensional system \cite{Dechant:2018:TUR}, molecular motors \cite{Hwang}, biochemical oscillations \cite{Bio}, interacting oscillators \cite{Interacting}, run-and-tumble process \cite{Mayank},
measurement and feedback control \cite{Van,SamuelssonM}, 
broken time reversal symmetry systems \cite{Udo:TURB,Saito,Garrahan18,SamuelssonM,Hyst}, 
first passage times \cite{Garrahan:2017:TUR,Passage} and quantum transport  problems
\cite{BijayTUR,TUR-bijay1,TURQ,SamuelssonQP,JunjieTUR, Goold}. 
Tighter bounds have also been reported for some stochastic currents \cite{Polettini:2016:TUP}. 

More recently, following the fundamental nonequilibrium fluctuation relation \cite{VanTUR}, 
a generalized version of the TUR (G-TUR1) was derived, where the RHS of Eq.~(\ref{eq:fund-TUR}) was modified to 
%
$ \frac{\langle Q^2 \rangle_c}{\langle Q \rangle^2} \geq \frac{2}{\exp{\langle \Sigma \rangle} -1}$,
%
which is a looser bound compared to Eq.~(\ref{eq:fund-TUR}). In fact, a more tighter version of the generalized bound had been obtained 
following a slightly different approach by Timpanaro {\it et al.} \cite{Landi-PRL} as
%
$ \frac{\langle Q^2 \rangle_c}{\langle Q \rangle^2} \geq f(\langle \Sigma \rangle)$,
%
where $f(x)={\rm csch}^2(g(x/2))$ and $g(x)$ is the inverse function of $x \tanh(x)$. 
We refer to this bound as the G-TUR2. Interestingly, in the small dissipation limit, $\langle \Sigma \rangle \to 0$, both 
these generalized bounds reduce to the 
S-TUR of Eq. (\ref{eq:fund-TUR}).

Despite intense theoretical efforts dedicated to derive and analyze the TUR, 
an experimental study of this trade-off relation is still missing.
In this work, we experimentally study the TUR of quantum heat exchange between two initially thermalized qubits in a NMR setup, 
in the transient regime. 
Moments of heat exchange are obtained by performing quantum state tomography (QST) for the qubits.
As expected,  G-TURs are valid throughout. This agreement, while fundamentally 
important, does not offer practical input for the design of quantum heat machines.
In contrast, by identifying violations of the S-TUR, observed in certain parameters and in excellent agreement with
analytical results, we can pinpoint favorable regimes of operation.


\vspace{2mm}
{\it Cumulants of heat exchange.}
%
Consider two systems with their Hamiltonians $H_1$ and $H_2$ that are initially  $(t<0)$ 
decoupled and separately prepared at their respective thermal equilibrium state. 
The initial composite density matrix is thus given as a product state,
${\rho}(0) = {\rho}_1 \otimes {\rho}_2$,  
with ${\rho}_{i} = \exp\big[{-\beta_{i} H_{i}}\big]/{\mathcal Z}_{i}, i=1,2$ the Gibbs thermal state 
with inverse temperature $\beta_{i}=1/k_B T_{i}$ ($k_B$ is the Boltzmann constant) and 
${\mathcal Z}_{i}={\rm Tr} \big[e^{-\beta_{i} H_{i}}\big]$ the corresponding equilibrium partition function. 
The coupling between the systems is suddenly switched on at
$t=0$ for a duration $\tau$ (total Hamiltonian $\mathcal{H}$), which allows energy exchange between the two systems.  
Due to the randomness of the initial thermal state and the inherent probabilistic nature of 
quantum mechanics, the exchanged energy is not a deterministic quantity, but rather 
quantified with a probability distribution function (PDF). 
In the quantum regime, this PDF is constructed by following a 
two-point projective measurement scheme \cite{fluct1, fluct2, campisi-measurement}: 
The first projective measurement of the energy of the two systems is performed before they are coupled.
A second  projective measurement is done at the end of the energy exchange process (after the systems are separated).
This  procedure respects the fundamental Jarzynski and W\"ojcik exchange fluctuation 
symmetry \cite{JarzW}. 
For the bipartite setup considered here, the joint PDF corresponding to energy change ($\Delta E_i, i=1,2$) between
the systems, during a coupling interval $\tau$ is denoted by $p_{\tau}(\Delta E_1, \Delta E_2)$. 
It can be shown that \cite{Wei,supp}
\bea
&&\Big \langle \big(e^{-\beta_1 \Delta E_1 \!-\!  \beta_2 \Delta E_2}\big)^z \Big \rangle_{\tau} 
\nonumber\\
&&= \! \int \!  \!d(\Delta E_1) d(\Delta E_2)  p_{\tau}(\Delta E_1,\Delta E_2) e^{-z \beta_1 \Delta E_1 \!-\!  z \beta_2 \Delta E_2} 
\nonumber\\
&&= {\rm Tr} \Big[ \rho(0)^z \, \rho(\tau)^{1-z}\Big],
\label{eq:central-Renyi}
\eea
with  $\rho(0)$ the combined density matrix of the two systems at the moment they are coupled, and $\rho(\tau)$
their density matrix at the end of their coupled evolution.
We now consider the case $\Delta E_1\approx -\Delta E_2$, 
which is justified when the two systems are only {\it weakly} coupled. Alternatively, this approximation becomes an exact equality if
there is no energy cost involved in turning on and off the interaction between the two systems. 
Interpreting the energy change for individual systems as heat, $\Delta E_1= - \Delta E_2 = Q$, 
we directly get from Eq. (\ref{eq:central-Renyi}) an expression for the moments of heat exchange \cite{supp}, 
\bea
\langle Q^n \rangle_{\tau} &=&\frac{1}{(\Delta \beta)^n}  {\rm Tr} \Big[ \rho(\tau){\rm T}_n \big(\ln \rho(\tau) \!-\! \ln \rho(0)\big)^n \Big],
\label{eq:expM}
\eea
where $n=1,2, \cdots$ corresponds to the order of the heat exchange moment and $\Delta \beta=\beta_1-\beta_2$.  ${\rm T}_n$ is the time-ordering operator; it places operators at the latest time to the left. 
This powerful expression offers a unique way to gather moments of heat exchange, simply by performing 
quantum state tomography based on NMR experiments.
Alternatively, cumulants of heat exchange can be obtained
by implementing an ancilla-based interferometric technique \cite{ancilla-1,ancilla-2,ancilla-3,Bijay-expt}. 
This method gives a direct access to the characteristic function (CF) of heat \cite{XFT-theory, XFT-agarwalla}, 
defined using the two-point measurement protocol,
\begin{eqnarray}
\!\!\!\!\!\!\chi_{\tau}(u)&\!\!=\!\!&\int dQ \, e^{i u Q} \, p_\tau(Q), \nonumber \\
=&&\!\!{\rm Tr}\Big[{\cal U}^{\dagger}(\tau,0) (e^{i u H_1} \otimes {1}_2) {\cal U}(\tau,0)  (e^{-i u H_1} \otimes {1}_2) \rho(0)\Big]. 
\nonumber 
\label{eq:CF-TTM}
\end{eqnarray}
Here $u$ is the variable conjugate to $Q$, 
${\cal U}(t,0)=e^{- i \mathcal{H} t/\hbar}$ is the unitary propagator with the total Hamiltonian $\mathcal{H}$.
In the language of the CF, the exchange fluctuation symmetry translates to
$\chi_{\tau}(u) = \chi_{\tau}\big(-u + i \Delta \beta \big)$ \cite{JarzW,Bijay12, Saito07,Lutz_2018, XFT-agarwalla}. 

\vspace{2mm}
{\it Theoretical analysis.}
We now describe a specific case, the so-called XY-model consisting two qubits with the Hamiltonian
\bea
\mathcal{H}_{XY} &=& \frac{h \nu_0}{2} \sigma_1^z \otimes 1_2 
+ 1_1 \otimes \frac{h \nu_0}{2} \sigma_2^z  
\nonumber\\
&+& \frac{h J}{2}\,( \sigma_1^x \otimes \sigma_2^y - \sigma_1^y \otimes \sigma_2^x).
\label{eq:htotal1}
\eea
%
Here, $H_1=\frac{h \nu_0}{2} \sigma_1^z \otimes 1_2$, $H_2= 1_1 \otimes \frac{h \nu_0}{2} \sigma_2^z$  
with $\nu_0$ the frequency of the qubits,
$\sigma_i, i=x,y,z$ are the standard Pauli matrices. 
The last term, denoted by $H_{12}$, represents the interaction between the qubits, with $J$ the coupling parameter.
An important feature of this model is that $[H_{12},H_1+H_2]=0$.
This commutation implies that the change of energy for one qubit is exactly compensated by the other qubit,
as there is no energy cost involved in turning on or off the interaction between the qubits. 
For such an `energy-preserving' Hamiltonian
$\Delta E_1 = -\Delta E_2 =Q$ is exact and  
the average entropy production simply reduces to $\langle \Sigma \rangle = (\beta_1-\beta_2)\, \langle Q\rangle$. 


 %
%
Cumulants of heat exchange 
can either be computed from the composite density matrix \cite{supp}, 
or directly from the CF $\chi_{\tau}(u)$ of heat, following Eq.~(\ref{eq:CF-TTM}). 
We take the latter approach for the XY-model;
algebraic manipulations of the Pauli matrices yield \cite{Bijay-expt}
%
\begin{eqnarray}
 \chi_{\tau}(u) &&=  \Big[1 + \sin^2 \Big(2 \pi J \tau\Big) \Big\{f_1(\nu_0) \, (1 - f_2(\nu_0)) \big(e^{-i h u \nu_0} -1\big) \nonumber \\
&&+ f_2 (\nu_0)(1 - f_1(\nu_0)) \big(e^{i h u \nu_0} -1\big)\Big\}\Big],
\label{eq:chi-ana}
\end{eqnarray}
where $f_{i}(\nu_0)= (e^{\beta_i h \nu_0} + 1)^{-1}$, $i=1,2$. For compactness, below
we identify these functions as $f_{1,2}$.
It is easy to verify that the above CF satisfies the exchange fluctuation symmetry
for arbitrary values of $J$, $\tau$, $\beta_1,\beta_2$, and $\nu_0$. 
Expressions for the average heat current and the associated noise are  derived
by taking successive derivatives of $\ln  \chi_{\tau}(u)$  with respect to $iu$.
We write down the first three cumulants, useful for the analysis of the TUR,
\begin{widetext}
\begin{eqnarray}
&&\langle Q \rangle_{\tau} = h \nu_0  {\cal T}_{\tau}(J) \Big[f_2 \!-\!f_1\Big], 
\nonumber\\
&&\langle Q^2 \rangle^{c}_{\tau}= (h \nu_0)^2  \Big[{\cal T}_{\tau}(J) \Big( f_1 (1\!-\!f_2) \!+\! f_2 (1\!-\!f_1)\Big)\!-\! {\cal T}_{\tau}^2(J) \,\big(f_2\!-\!f_1\big)^2 \Big],\,\,\,\,\,\,  
\nonumber\\
&&\langle Q^3 \rangle^{c}_{\tau}=  (h \nu_0)^3  \, {\cal T}_{\tau}(J) (f_1\!-\!f_2) \Big[ 1 \!-\! 3\, {\cal T}_{\tau}(J) \big(f_1(1\!-\!f_2) + (1\!-\!f_1) f_2 \big) + \, 2 \, {\cal T}_{\tau}^2(J) (f_1-f_2)^2 \Big].
\label{eq:analM}
\end{eqnarray}
\end{widetext}
Here, ${\cal T}_{\tau}(J) = \sin^2 \Big(2 \pi J \tau\Big)$. 
 


\hspace{2mm}
{\it Perturbative expansion of the S-TUR.}
For arbitrary coupling time $\tau$,
the cumulants can be expanded close to equilibrium in terms of the thermal affinity $\Delta \beta = \beta_1 -\beta_2$, 
around a fixed inverse temperature $\beta$. Specifically, 
\begin{eqnarray}
\langle Q \rangle_{\tau} &=& G_1(\tau) \Delta \beta +  G_2(\tau) \frac{(\Delta \beta)^2}{2!}  + G_3(\tau) \frac{(\Delta \beta)^3}{3!}  + \cdots \nonumber \\
\langle Q^2 \rangle^c_{\tau} &=& S_0(\tau) + S_1(\tau) \Delta \beta + S_2(\tau) \frac{(\Delta \beta)^2}{2!}  + \cdots \nonumber \\
\langle Q^3 \rangle^c_{\tau} &=& R_1(\tau) \Delta \beta +  \cdots 
\end{eqnarray}
Here $G_1(\tau)$ is the time-dependent linear transport coefficient and $S_0(\tau)$ is the equilibrium 
noise. $G_2(\tau)$, $G_3(\tau), \cdots $ ($S_1(\tau)$, $S_2(\tau), \cdots$) are higher order nonequilibrium transport (noise) coefficients.  
As a consequence of the exact fluctuation symmetry, the following relations hold \cite{SaitoUts}:
$S_0(\tau) = 2 G_1(\tau)$,  $S_1(\tau) = G_2(\tau)$,
$3 S_2 (\tau) - 2 G_3(\tau) = R_1(\tau) $, and so on.
This leads to \cite{TUR-bijay1},
\begin{equation}
\langle \Sigma \rangle \frac{\langle Q^2 \rangle^c_{\tau}}{\langle Q \rangle_{\tau}^2} = \Delta \beta \frac{\langle Q^2 \rangle^c_{\tau}}{\langle Q \rangle_{\tau}}= 2 + \frac{(\Delta \beta)^2}{6} \frac{R_1(\tau)}{G_1 (\tau)} + {\cal O}(\Delta \beta)^3.
\label{eq:STURf}
\end{equation}
Interestingly, the contribution of the linear term $\Delta \beta$ disappears;
the presence of this term could trivially violate the S-TUR by swapping
the initial temperatures of the qubits.
%
%
While the linear coefficient for the average heat exchange, $G_1(\tau)$, 
is always positive,  $R_1(\tau)$ does not take a definite sign; when
$R_1(\tau)<0$, the S-TUR is violated. 
For the XY-model we get  ($f(\nu_0)$ is evaluated at $\beta$),
\begin{eqnarray}
G_1(\tau) &=& (h \nu_0)^2 \, {\cal T}_{\tau}(J) \, f (1\!-\!f) \geq 0, \nonumber \\
R_1(\tau) &=& (h \nu_0)^4 \, {\cal T}_{\tau}(J) \, f (1\!-\!f) \Big[1\!-\!6 {\cal T}_{\tau}(J) f (1\!-\!f) \Big]
\end{eqnarray}
To order $(\Delta \beta)^2$, Eq. (\ref{eq:STURf}) simplifies to 
\bea
 \Delta \beta \frac{\langle Q^2 \rangle^c_{\tau}}{\langle Q \rangle_{\tau}}= 2 +(\Delta \beta h\nu_0)^2 \Big[\frac{1}{6}- {\cal T}_{\tau}(J) f(1-f)\Big].
 \label{eq:TURorder2}
\eea
The S-TUR is violated when $R_1(\tau)<0$, that is
${\cal T}_{\tau}(J) f\,(1-f) > 1/6$. However, since $0 \leq f(1-f) \leq 1/4$, the S-TUR is violated 
once ${\cal T}_{\tau}(J) > \frac{2}{3}$.
Interestingly, already in the quadratic order of $\Delta \beta$ the TUR can drop below the value of 2 if 
${\cal T}_{\tau}(J)$  crosses a critical value. 
We assess the perturbative formula (\ref{eq:TURorder2}) in Ref. \cite{supp}.
%
However, in the weak coupling limit i.e., $J\tau \ll 1$, ${\cal T}_{\tau}^2(J) \ll {\cal T}_{\tau}(J)$,
and $R_1(\tau)$ is always positive. 
Moreover, it can be shown that in this limit the S-TUR bound is always above 2, even far from equilibrium 
\cite{comment}.
%

\begin{figure}
\includegraphics[trim= 0.4cm 0cm 0cm 0cm, clip=true,width=1.05\columnwidth]{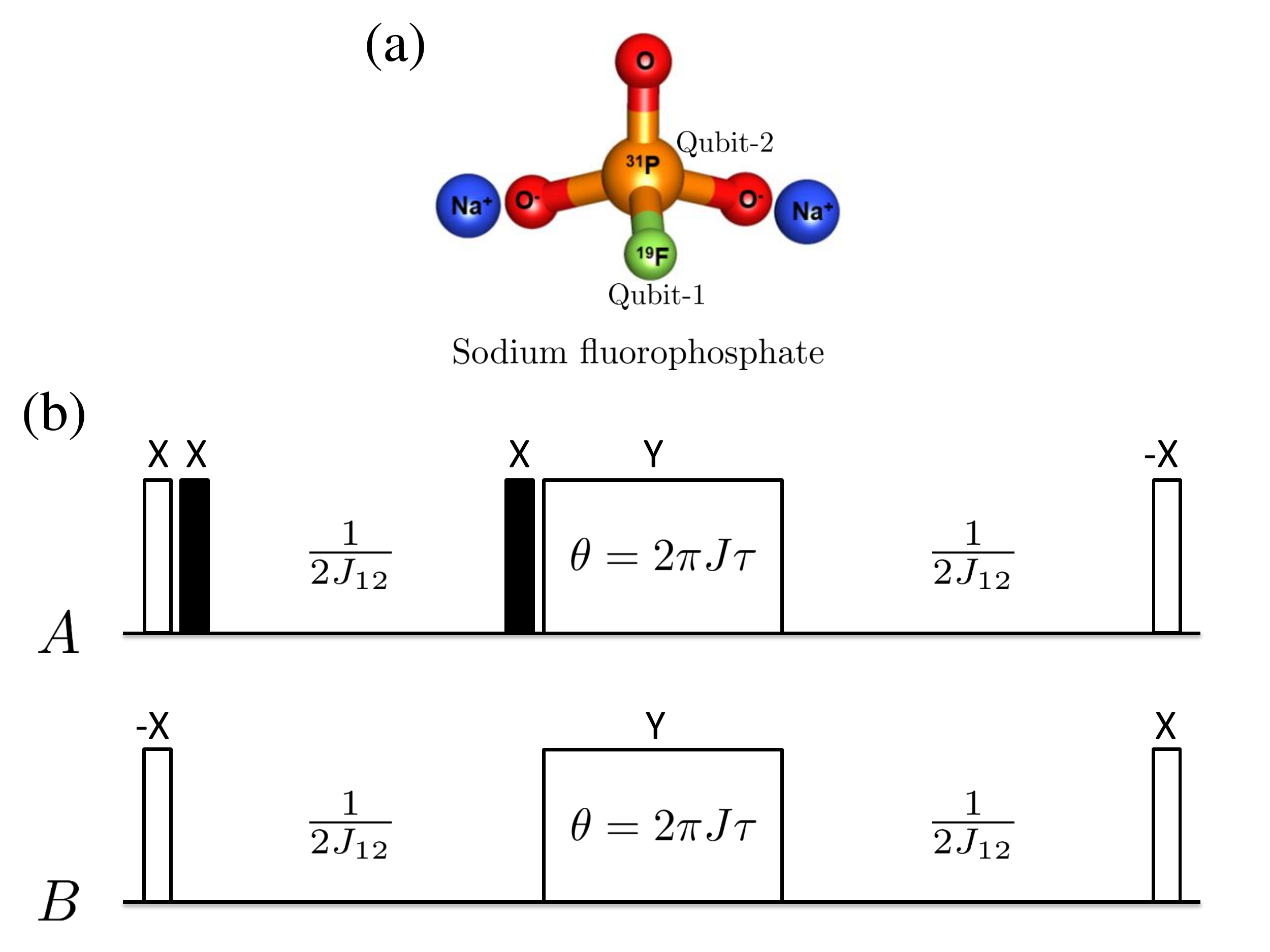} 
\caption{
(a) Molecular structure of the two-qubit NMR spin system, 
Sodium fluorophosphate. The NMR active spin-1/2, $^{19}$F and $^{31}$P nuclei in the molecule, 
labeled as qubit $1$ and qubit $2$ respectively, are coupled by the Hamiltonian (\ref{eq:intH})
with the coupling strength $J_{12} = 868$\,Hz. 
(b) Pulse sequence to realize heat exchange coupling Hamiltonian, 
$\mathcal{H}_{XY}$ in Eq. (\ref{eq:htotal1}). The pulses are applied on qubits  
$1$ and $2$ in a time ordered manner from left to right. 
The black and white narrow solid bars represent $\pi$ and $\pi/2$ pulses, respectively, 
with the phases mentioned above them.  $1/2J_{12}$ represents the free evolution delay. 
The white box represents the $\theta$ (in rads) angle pulse about y-axis.} 
\label{molecule_pulse}
\end{figure}

\begin{figure}
\includegraphics[trim= 1cm 0cm 0cm 0cm, clip=true,width=1.05\columnwidth]{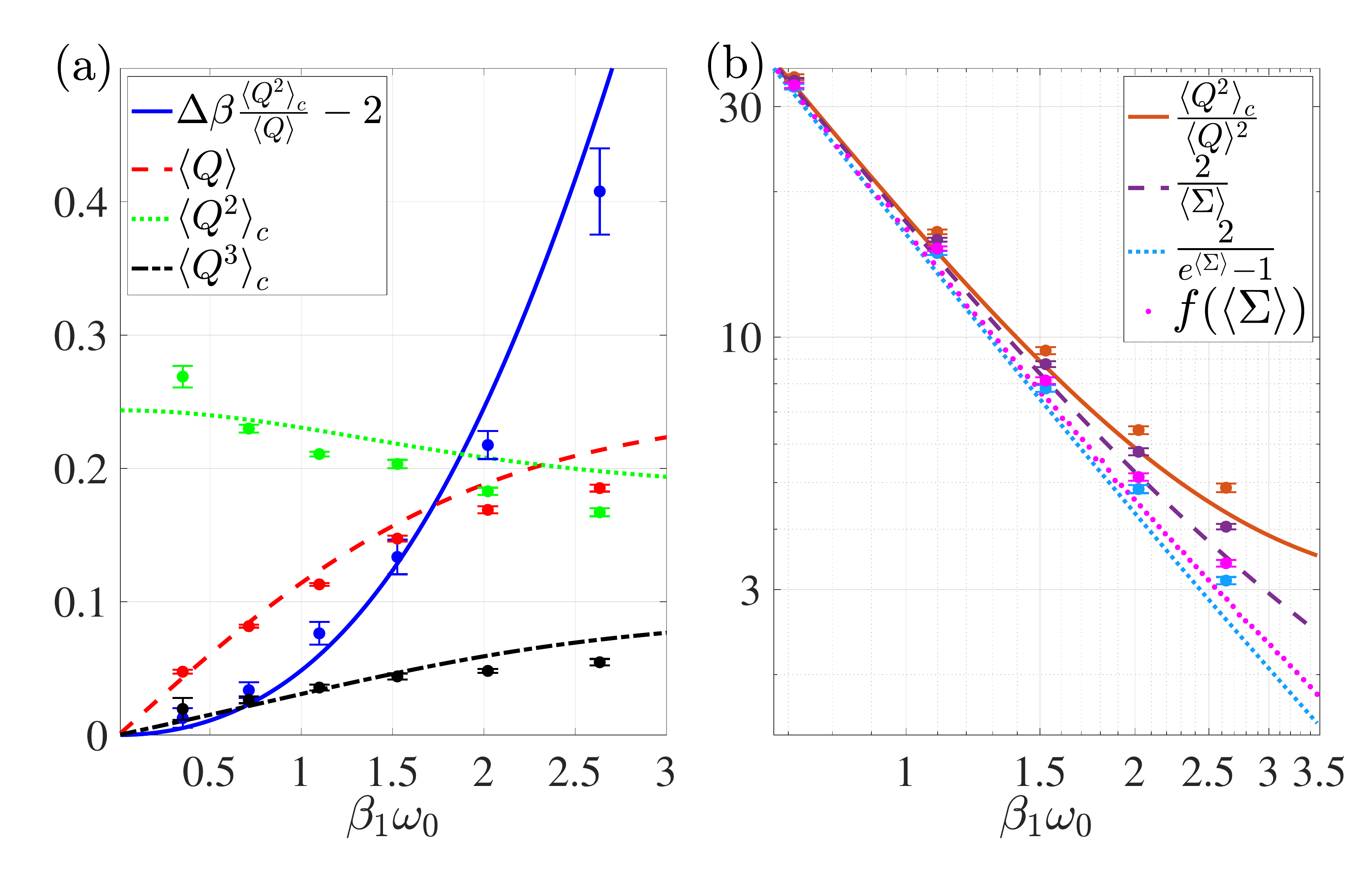} 
\caption{
(a) First three cumulants of heat exchange, along with a measure for the S-TUR,
as a function of the inverse temperature of qubit 1 $\beta_1$; $\beta_2=0$.
Measurements (symbols) are constructed with the help of Eq. (\ref{eq:expM}), and are compared to
the theory (lines), Eq. (\ref{eq:analM}).
(b) Comparison between different bounds, showing that the S-TUR provides the tightest lower bound to 
$\frac{\langle Q^2\rangle^c_{\tau}}{\langle Q \rangle_{\tau}^2}$.
Experimental results are obtained from state tomography, yielding $\langle Q\rangle_{\tau}$, which is used to
calculate the entropy production. Theoretical results are based on Eq. (\ref{eq:analM}).
Parameters are $J\tau = 1/8$ and $\nu_0 = \pi/20$ ($\omega_0 = 2\pi\nu_0$).
Error bars are obtained by repeating the experiments 8 times.} 
\label{TUR_NoVio}
\end{figure}



\begin{figure}
\includegraphics[trim= 0.6cm 0cm 0cm 0cm, clip=true,width=1.05\columnwidth]{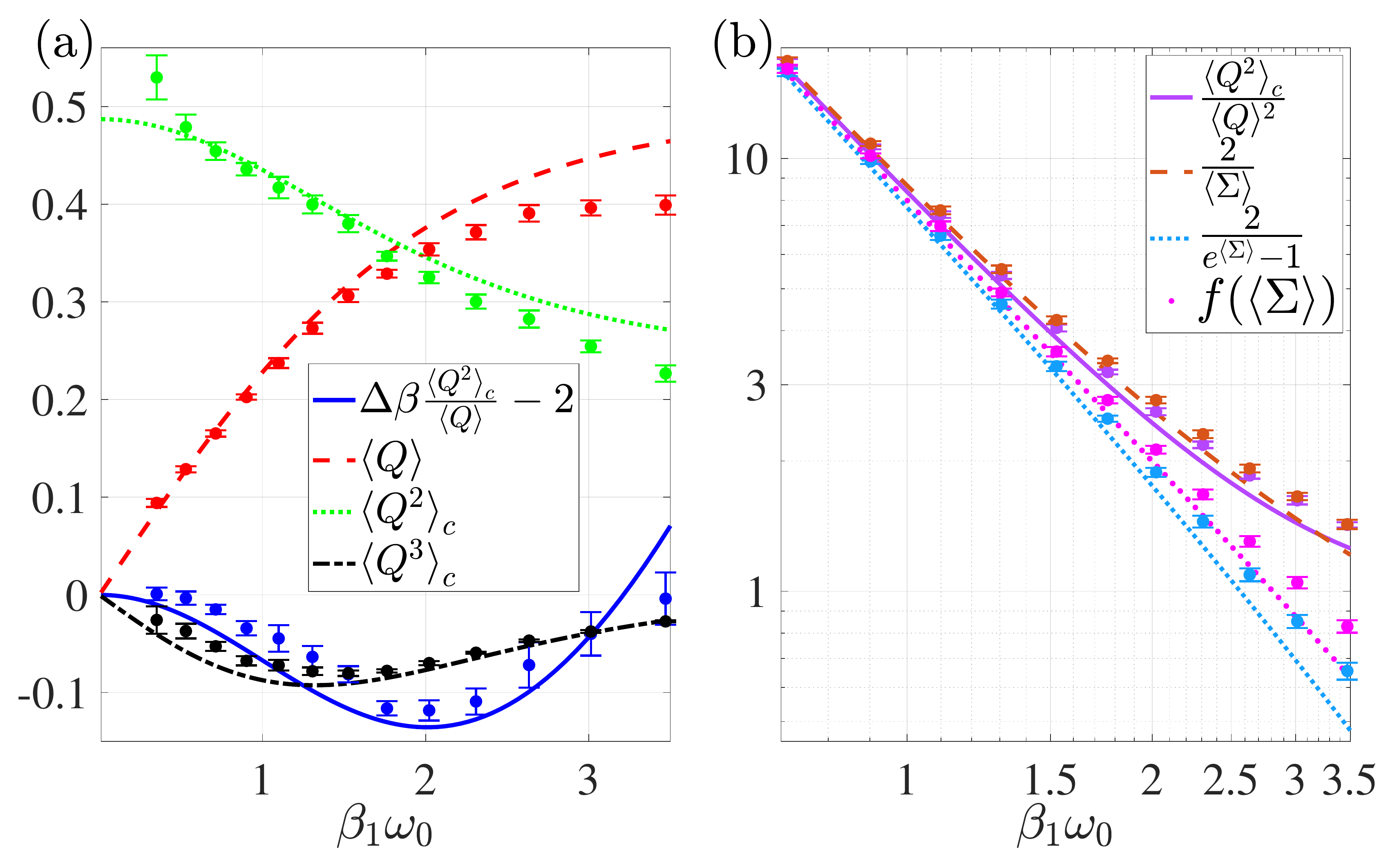} 
\caption{Same as Fig. \ref{TUR_NoVio} but at $J\tau=1/4$
leading to $\mathcal T_{\tau}(J)>2/3$, therefore the violation of the S-TUR.}
\label{TUR_Vio}
\end{figure}

\begin{figure}
\includegraphics[trim=1cm 1cm 1cm 1cm, clip=true,width=0.85\columnwidth]{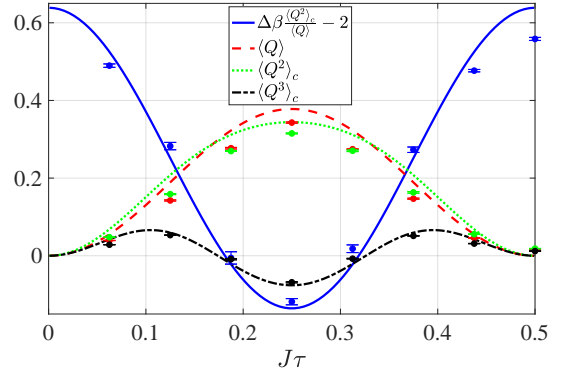} 
\caption{Cumulants of heat exchange and the S-TUR
as a function of $J\tau$  for $J=1$ {\rm Hz}, $\beta_1 \omega_0= 2.02$  
and $\beta_2=0$. Other parameters are the same as in Fig.~(\ref{TUR_NoVio}).
}
\label{Jtau}
\end{figure} %

\vspace{2mm}
{\it Experimental setup and Results.}
To study heat exchange between two qubits
we use liquid-state NMR spectroscopy of the $^{19}F$ and $^{31}P$ 
nuclei in the molecule Sodium fluorophosphate dissolved in D$_2$O. 
Experiments are performed in 500MHz Bruker NMR spectrometer at ambient temperature. 
As shown in Fig. \ref{molecule_pulse}(a), $^{19}$F and $^{31}$P are identified as the two qubits, 1 and 2, 
exchanging heat under the desired coupling Hamiltonian, Eq. (\ref{eq:htotal1}). 
As the sample is in the liquid state, the molecules can be considered identical with intermolecular interactions 
averaged out due to motional averaging. 
All the experimental procedures: initialization of the system and the heat-exchange, 
are completed in time scales much shorter  than the relaxation time of the nuclei. 
The internal Hamiltonian $H_{\rm{int}}$ of the two spins---in the rotating frame of the radio frequency (RF) pulses---can be written as
%
\begin{equation}
H_{\rm int} =  \frac{\pi}{2} J_{12} \sigma^z_1  \sigma^z_2,
\label{eq:intH}
\end{equation}
where $J_{12} = 868$ Hz is the scalar coupling between the
$^{19}$F and $^{31}$P nuclei, as explained in Fig. \ref{molecule_pulse}(a). 
The desired coupling Hamiltonian, $\mathcal{H}_{XY}$, under which the spins exchange heat 
 is realized from the internal Hamiltonian $H_{\rm int}$
with the RF pulses displayed in Fig. \ref{molecule_pulse}(b). 
The net effect of the pulse sequence is that the two spins evolve under the coupling Hamiltonian 
$\mathcal{H}_{XY}$ for a duration $\tau$ that is specified by the $\theta$ angle rotation about y-axis, as shown. 
For the duration of $1/2J_{12}$, the system evolves under the Hamiltonian $H_{\rm int}$. 

To start with, the two qubits are initialized in a psuedoequilibrium state 
$\rho_1 \otimes \rho_2$, where $\rho_i = \exp\big[{-\beta_i H_i}\big]/ {\cal Z}_i $ 
is a Gibbs thermal state with inverse 
pseudo spin temperatures $\beta_i$ and ${\cal Z}_i$ the partition function. 
For simplicity, we set $\beta_2 = 0$ in all our measurements. 
Qubit $1$ is prepared at a higher inverse temperature $\beta_1$ by initializing it in a pseudopure 
state (PPS) of $\ket{0}\langle0|$, followed by applying pulses between 0 and $\pi/2$, 
and a pulse field gradient (PFG). The purpose of the PFG is to destroy coherences produced by 0 to $\pi/2$ angle pulses. 
The qubits---prepared at two different pseudoequilibrium states---are made to exchange heat under the coupling Hamiltonian $\mathcal{H}_{XY}$ for different time interval 
$\tau$ and different $\beta_1$. 
Following the coupling period, we perform QST of the final state 
(in addition to the QST of the initial pseudoequilibrium state) \cite{supp},
and from Eq.~(\ref{eq:expM}) achieve the cumulants of heat exchange.

In Figs. \ref{TUR_NoVio} and \ref{TUR_Vio} we present
two cases, 
displaying agreement and violation, respectively, of the S-TUR.
First, in Fig. \ref{TUR_NoVio} we set $J\tau=1/8$.
According to the theoretical analysis, the S-TUR is valid 
when the skewness is positive, or ${\cal T}_{\tau}(J)=1/2<2/3$.
Indeed, we find in Fig. \ref{TUR_NoVio}(a) that both $R_1(\tau)$ 
and $\Delta \beta \frac{\langle Q^2 \rangle^c_{\tau}}{\langle Q \rangle_{\tau}}-2$ are positive for all $\Delta \beta$.  
In Fig. \ref{TUR_NoVio}(b), 
we compare the different bounds on the relative uncertainty
$\frac{\langle Q^2 \rangle^c_{\tau}}{\langle Q \rangle_{\tau}^2}$, using experimental data
as well as theoretically, and show that the S-TUR provides the tightest bound. 
%
%
Next, in Fig. \ref{TUR_Vio}(a) we display results for $J\tau = 1/4$, for which according to our theory violations of the S-TUR are expected to occur already in the quadratic order of $\Delta \beta$, as ${\cal T}_{\tau}(J)=1>2/3$ . 
Indeed, we clearly see a violation for  $0<\beta_1\omega_0<3.2$. 
Furthermore, the third cumulant, $\langle Q^3 \rangle_c$,
is negative in this region, which corroborates with Eq. (\ref{eq:TURorder2}). 
The theoretically predicted lowest value for the S-TUR for this model is 
$\Delta \beta \frac{\langle Q^2\rangle^c_{\tau}}{\langle Q\rangle_{\tau}}\approx 1.86$, and
we experimentally reach a value very close to this number.
The violation of the S-TUR can also be seen in Fig. \ref{TUR_Vio}(b): 
The  S-TUR bound ($2/\langle \Sigma \rangle$) appears {\it above} 
the ratio   $\frac{\langle Q^2\rangle^c_{\tau}}{\langle Q\rangle_{\tau}^2}$,
and it is greater than the other, looser bounds. 
Measurements again closely match the theoretical curves. 

A complete analysis of the TUR as a function of the heat exchange duration $\tau$ 
and for a fixed $J=1$ Hz,
is presented in Fig. \ref{Jtau}. We display the first three cumulants 
and note that the relative uncertainty is reduced (violation of S-TUR) within a certain region of parameters: The minimum value of the S-TUR precisely appears when the fluctuation of the heat exchange are
reduced, below the value of the first cumulant. As expected, the skewness is found to be negative in this region.  

{\it Summary.}
We experimentally examined the TUR for heat exchange by realizing the XY-model, 
performing quantum state tomography and extracting the heat exchange cumulants. 
We found that the S-TUR provides a tight bound 
up to a certain threshold value for the qubit-qubit coupling parameter $\sin ^2 (2\pi J\tau)$,
beyond which the bound is invalidated. As predicted theoretically, 
the validity of the S-TUR crucially depends on the sign of the third cumulant.
Generalized versions of the TUR are  satisfied throughout, as expected, since these (loose) bounds are derived 
from the universal fluctuation relations. Nevertheless, 
the S-TUR contains more information: The condition to invalidate it pinpoints to
regimes of favorable performance for heat machines, operating with high constancy {\it and} little dissipation.

{\bf Acknowledgment}
BKA gratefully acknowledges the start-up funding from IISER Pune and the Max Planck-India mobility grant. 
TSM acknowledges the support from the Department of Science and Technology, India (Grant
Number DST/SJF/PSA-03/2012-13) and the Council of Scientific and Industrial Research (CSIR), India (Grant Number CSIR-03(1345)/16/EMR-II). SS acknowledge support from CSIR, India (Grant Number 1061651988). 
The work of DS is supported by the Canada Research Chairs program and an NSERC discovery grant.
BKA thanks Gabriel Landi and Gernot Schaller, SS thanks Deepak Dhar and SP thanks V. R. Krithika for insightful discussions.


\newpage

\renewcommand*{\thepage}{S\arabic{page}}
\renewcommand*{\thefigure}{S\arabic{figure}}

\setcounter{equation}{0}  
\setcounter{section}{0}
\setcounter{table}{0}
\setcounter{figure}{0}
\renewcommand{\thesection}{S\arabic{section}}
\renewcommand{\theequation}{S\arabic{equation}}

\begin{widetext}
\begin{center}
{\huge Supplemental Material}
\end{center}

\section{Relation between energy exchange and R\'enyi divergences}

We provide here details on the derivation of Eq. (\ref{eq:central-Renyi}), which generalizes results of Ref. \cite{Wei}.
We consider two systems with Hamiltonians $H_1$ and $H_2$, initially decoupled and prepared at their respective thermal equilibrium state. 
The initial composite density matrix is a product state, ${\rho}(0) = {\rho}_1 \otimes {\rho}_2$  
with ${\rho}_{i} = \exp\big[{-\beta_{i} H_{i}}\big]/{\mathcal Z}_{i}, i=1,2$ a Gibbs thermal state with 
inverse temperature $\beta_{i}=1/k_B T_{i}$; $k_B$ the Boltzmann constant.
${\mathcal Z}_{i}={\rm Tr} \big[\exp(-\beta_{i} H_{i})\big]$  is the corresponding equilibrium partition function. 
The systems are coupled at $t=0$ for a period $\tau$,
which allows energy exchange between the two systems. 
This exchange of energy is not a deterministic process, but it is described by a probability distribution function (PDF). 
In the quantum regime, the PDF of energy exchange is constructed from a two-point projective measurement 
protocol \cite{fluct1, fluct2, campisi-measurement} performed at the beginning of the energy exchange process, and after decoupling.
This procedure respects the Jarzynski and W\"ojcik exchange fluctuation symmetry \cite{JarzW}. 
For bipartite setups, we construct the joint PDF corresponding to energy change 
($\Delta E_i, i=1,2$) for both the systems, given as
\begin{equation}
p_{\tau}(\Delta E_1, \Delta E_2)\!=\! \sum_{m,n} \Big(\prod_{i=1}^2 \,\delta(\Delta E_i - (\epsilon_m^i -\epsilon_n^i)) \Big) p_{m|n}^\tau p_{n}^{0}.
\end{equation}
Here, $p_n^0 = \prod_{i=1}^2 e^{-\beta_i \epsilon_n^i}/{\mathcal Z}_i$ 
is the probability to find the decoupled systems in the 
eigenstate $|n \rangle = |n_1,n_2\rangle$ with energy eigenvalues $\epsilon_n^i$, 
$H_i|n_i\rangle=\epsilon_n^i|n_i\rangle$, after the first projective measurement. 
The second projective measurement at $t=\tau$ collapses the system to the eigenstate $|m\rangle= |m_1,m_2\rangle$,
$H_i|m_i\rangle=\epsilon_m^i|m_i\rangle$.
The corresponding transition probability is $p_{m|n}^\tau = \langle m |{\cal U}(\tau,0)|n\rangle|^2$, where
 ${\cal U}(t,0)=e^{- i \mathcal{H} t/\hbar}$ is the unitary propagator with the total-composite 
Hamiltonian $\mathcal{H}$. 
The principle of microreversibility of quantum dynamics for autonomous systems demands 
$p_{m|n}^\tau =p_{n|m}^\tau$. 
Following this relation and given the uncorrelated initial thermal condition for the composite system, 
we receive the following universal symmetry for the joint PDF,
\begin{equation}
p_{\tau}(\Delta E_1,\Delta E_2) = e^{\beta_1 \Delta E_1 +  \beta_2 \Delta E_2} \,p_{\tau}(-\Delta E_1, -\Delta E_2).
\label{eq:Sfluc}
\end{equation}
This symmetry motivates us to define a characteristic function-like quantity, which leads to the following crucial relation
\begin{eqnarray}
\Big \langle \big(e^{-\beta_1 \Delta E_1 \!-\!  \beta_2 \Delta E_2}\big)^z \Big \rangle_{\tau} \!&\equiv& 
\! \int \!  \!d(\Delta E_1) d(\Delta E_2)  p_{\tau}(\Delta E_1,\Delta E_2)
 e^{-z \beta_1 \Delta E_1 \!-\!  z \beta_2 \Delta E_2} 
\nonumber\\
&=& {\rm Tr} \Big[ \rho(0)^z \, \rho(\tau)^{1-z}\Big]. 
\nonumber\\
&=& \exp\Big\{(z-1) S_z\Big[\rho(0)||\rho(\tau)\Big]\Big\}.
\label{eq:Scentral-Renyi}
\end{eqnarray}
Here $S_z\Big[\rho(0)||\rho(\tau)\Big]\equiv \frac{1}{z-1} {\rm ln} \big\{{\rm Tr}\big[\rho(0)^z \, \rho(\tau)^{1-z}\big]\big\}$ 
is the order-z Renyi divergence, a metric for the relation between the 
states of a composite system at the initial ($t=0$) and the final ($t=\tau$) times.
As a special case, when $z=1$ we receive the universal relation, 
$\Big \langle e^{-\beta_1 \Delta E_1 \!-\!  \beta_2 \Delta E_2} \Big \rangle=1$. 

So far, the analysis is exact. However, it is relevant to consider the limit $\Delta E_1 \approx -\Delta E_2$, 
which is justified when the two systems are  weakly coupled. Furthermore,
$\Delta E_1 = -\Delta E_2$ if there is no energy cost involved in turning on and off the interaction between
the systems. 
One can then interpret the energy change for an individual system as heat ($\Delta E_1= - \Delta E_2 = Q$).
This modifies the symmetry relation in Eq. (\ref{eq:Sfluc}) for the joint PDF to 
$p_{\tau}(Q)= \exp\big[(\beta_1 -\beta_2) Q\big] p_{\tau}(-Q)$. Accordingly,
Eq.~(\ref{eq:Scentral-Renyi}) leads to
\begin{equation}
\langle \big(e^{-\Delta \beta Q}\big)^z \rangle_{\tau} =\exp\Big\{(z-1) S_z\Big[\rho(0)||\rho(\tau)\Big]\big\},
\label{eq:Sheat-Renyi}
\end{equation}
which immediately generates expressions for the moments, 
\begin{eqnarray}
\langle Q^n \rangle_{\tau} &=&\frac{1}{(\Delta \beta)^n}  {\rm Tr} \Big[ \rho(\tau){\rm T}_n \big(\ln \rho(\tau) \!-\! \ln \rho(0)\big)^n \Big].
\label{eq:Scumulant}
\end{eqnarray}
${\rm T}_n$ is the time-ordering operator, which orders operators at the latest time to the left and $\Delta \beta = \beta_1-\beta_2$.

\section{Derivation for heat exchange cumulants from the composite density matrix}

The initial density matrix for the composite system is given by a direct product of the individual qubits, each prepared in an equilibrium state with a particular temperature. In the matrix form, we can then write,
\begin{eqnarray}
\rho(0) = 
\begin{bmatrix}
f_1 f_2  & 0  & 0 & 0 \\
 0  & f_1 (1-f_2)   &0 & 0 \\
0 &0 & f_2 (1-f_1) & 0 \\
0 & 0 & 0 & (1-f_1) (1-f_2) \\
\end{bmatrix}. \nonumber
\end{eqnarray}
where $f_i(\nu_0)= 1/\big(\exp(\beta_i h \nu_0) +1\big)$.
The density matrix evolves under the interaction Hamiltonian according to the Liouville equation  
  $\rho(\tau)= {\cal U}(\tau,0) \, \rho(0)\, {\cal U}^{\dagger}(\tau,0)$ where ${\cal U}(t,0)= e^{- i \mathcal{H} t/\hbar}$ and for the XY model is given by,
\begin{eqnarray}
{\cal U}(\tau,0) = 
\begin{bmatrix}
e^{-2 i \pi \tau \nu_0} & 0  & 0  & 0  \\
 0  & \cos(2 \pi J\tau)   & \sin(2 \pi J\tau) & 0 \\
0 & -\sin(2 \pi J\tau) & \cos(2 \pi J\tau) & 0 \\
0 & 0 & 0 & e^{2 i \pi \tau \nu_0} \\
\end{bmatrix}. \nonumber
\end{eqnarray}
The density matrix for the composite system at any arbitrary heat exchange duration time $\tau$ can be analytically found, and is given as
\begin{eqnarray}
\rho(\tau) = 
\begin{bmatrix}
f_1 f_2  & 0  & 0 & 0 \\
 0  & f_1 (1-f_2) \cos^2(2\pi J \tau) + f_2 (1-f_1) \sin^2(2\pi J \tau)   & \frac{1}{2} \sin(4 \pi J\tau) (f_2 - f_1) & 0 \\
0 & \frac{1}{2} \sin(4 \pi J\tau) (f_2 - f_1) & f_2 (1-f_1) \cos^2(2 \pi J \tau) + f_1 (1-f_2) \sin^2(2 \pi J \tau) & 0 \\
0 & 0 & 0 & (1-f_1) (1-f_2) \\
\end{bmatrix}. \nonumber
\end{eqnarray}
One can similarly find the logarithm of this matrix,
\begin{eqnarray}
{\log \rho(\tau)} =
\begin{bmatrix}
\log(f_1 f_2)  & 0  & 0 & 0 \\
0 &  \log\big[f_1(1-f_2)\big] + \Delta \beta h\nu_0  \sin^2(2 \pi J \tau) & \frac{1}{2} \Delta \beta h \nu_0 \sin(4 \pi J\tau) & 0 \\
0 &\frac{1}{2} \Delta \beta h  \nu_0 \sin(4 \pi J\tau) &  \log\big[f_2(1-f_1)\big] -\Delta \beta h \nu_0  \sin^2(2 \pi J \tau) & 0 \\
0 & 0 & 0 & \log\big[(1-f_1)(1-f_2)\big] \\
\end{bmatrix}. \nonumber
\end{eqnarray}
We substitute these expressions for the composite density matrix into Eq. (\ref{eq:Scumulant}) and receive all moments for heat exchange.

\begin{figure}[htb]
\includegraphics[width=0.7\columnwidth]{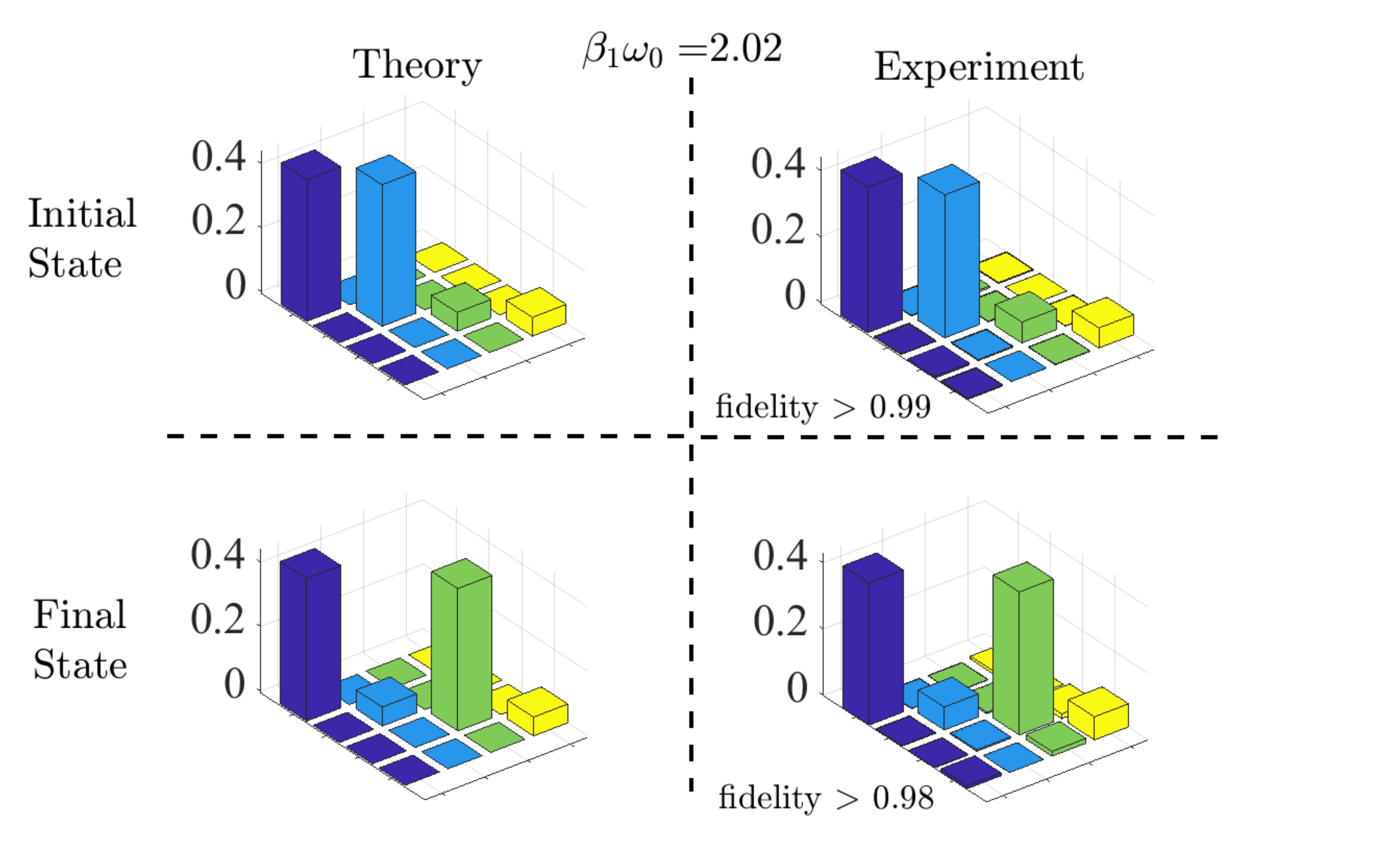}
\caption{Quantum state tomography for the real components of the density matrix elements for
 both initial and final states. Parameters are $J\tau=1/4$, $\beta_2=0$, $\nu_0=\pi/20$, $\beta_1 \omega_0=2.02$,
corresponding to Fig. \ref{TUR_Vio}.
}
\label{tomo}
\end{figure}

\section{Quantum state tomography of the XY model}
In Fig.~(\ref{tomo}) we provide both theoretical and the experimental quantum state tomography results for a particular realization. We display only the real components for both the initial and final density matrices of the composite system. The imaginary components for both these states are vanishingly small. In our tomography experiments the states are realized with fidelity higher than 97\%.    Both the initial and final states are obtained by performing 6 independent experiments and measurements.

\section{Validity of the perturbative expansion}
For the XY model, 
the ratio $\Delta \beta \frac{\langle Q^2\rangle^c_{\tau}}{\langle Q\rangle_{\tau}}$  
can be simulated exactly using the closed-form expressions for the cumulants, Eq. (\ref{eq:analM}).
Thoroughout the paper, these exact expressions were used to compare with measurements.
Nevertheless, the $(\Delta \beta)^2$ perturabtive analysis of the S-TUR, Eq. (\ref{eq:TURorder2}), is constructive
as it serves to quickly identify S-TUR violations:
The S-TUR is disobeyed if ${\mathcal T}_{\tau}(J)>2/3$.
In Fig. \ref{figS} we display the ratio
 $\Delta \beta \frac{\langle Q^2\rangle^c_{\tau}}{\langle Q\rangle_{\tau}}-2$ based on the exact expressions 
for the cumulants, and compare is to $C_2(\Delta\beta)\equiv \left[ \frac{1}{6}- {\cal T}_{\tau}(J) f(1-f) \right]$,
which measures the deviation from the equilibrium value, as received in Eq. (\ref{eq:TURorder2}).
We observe an excellent agreement up to  $\beta_1\omega_0\approx 1$, and meaningful results
up to $\beta_1\omega_0\approx 1.5$.  For larger $\Delta \beta$, the quadratic expansion obviously fails to
track the recovery of the S-TUR in Fig. \ref{figS} (b).

\begin{figure}[htpb]
\includegraphics[width=0.7\columnwidth]{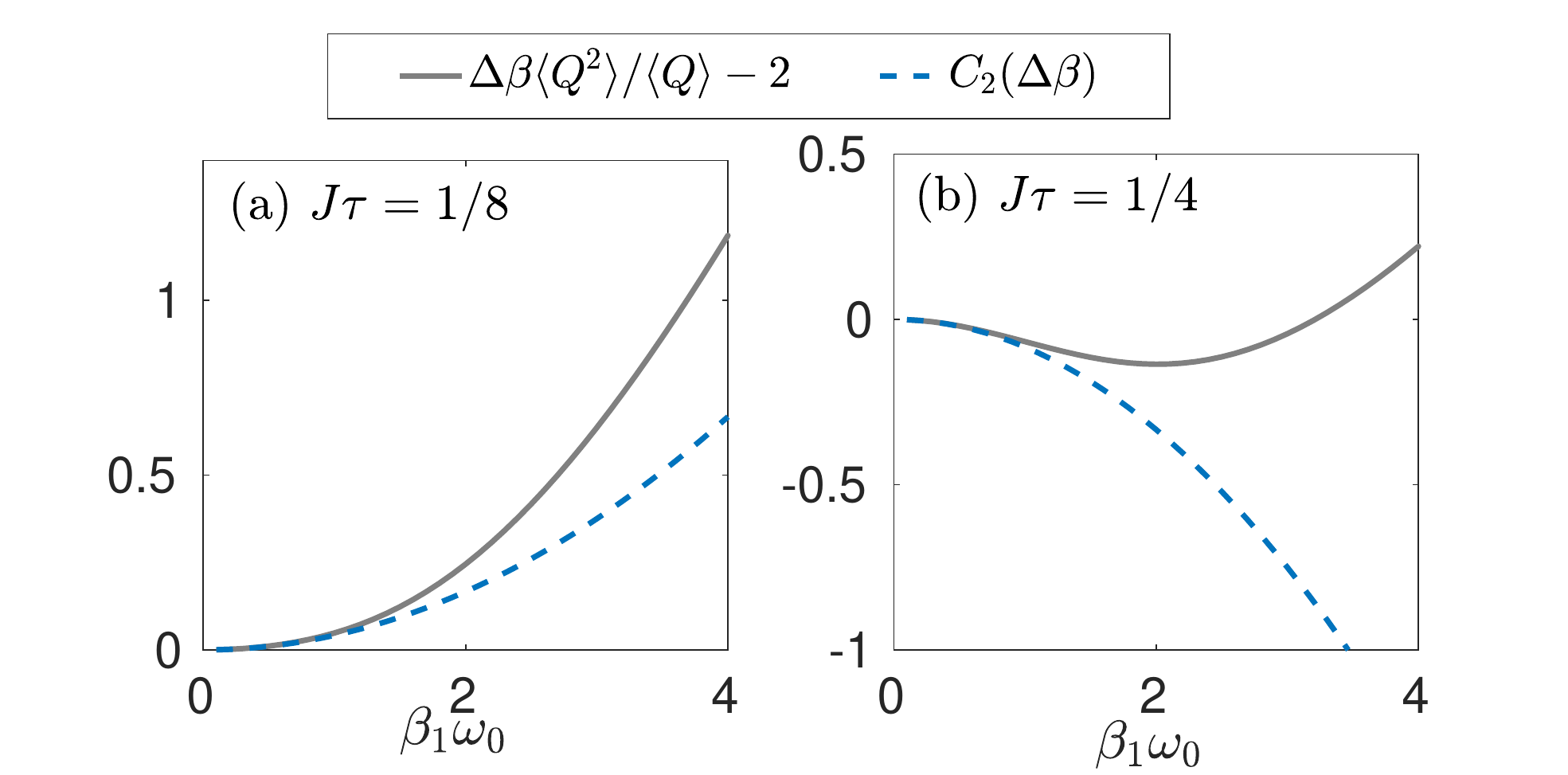} 
\caption{Analysis of the S-TUR based on exact expressions for the cumulants (full) and
the $(\Delta\beta )^2$ expansion (dashed), see text for the defintion of $C_2(\Delta \beta)$.
	(a) $J\tau=1/8$ thus ${\mathcal T}_{\tau}(J)<2/3$, corresponding to Fig. \ref{TUR_NoVio}.
(b) $J\tau=1/4$ thus ${\mathcal T}_{\tau}(J)>2/3$, corresponding to Fig. \ref{TUR_Vio}.
Parameters are $\beta_2=0$ and $h\nu_0=1$.
}
\label{figS}
\end{figure}

\end{widetext}


\begin{thebibliography}{999}

\bibitem{Barato:2015:UncRel}
A. C. Barato and U. Seifert,
Thermodynamic uncertainty relation for biomolecular processes,
Phys. Rev. Lett. {\bf 114}, 158101 (2015).

\bibitem{Gingrich:2016:TUP}
T. R. Gingrich, J. M.  Horowitz, N. Perunov, and J. L. England,
Dissipation bounds all steady state current fluctuations,
Phys. Rev. Lett. {\bf 116}, 120601 (2016).


\bibitem{Polettini:2016:TUP}
M. Polettini, A. Lazarescu, and M. Esposito, 
Tightening the uncertainty principle for stochastic currents, 
Phys. Rev. E {\bf 94}, 052104 (2016).

\bibitem{Pietzonka:2016:Bound}
P. Pietzonka, A. C. Barato, and U. Seifert, 
Universal bounds on current fluctuations,
Phys. Rev. E {\bf 93}, 052145 (2016).

\bibitem{Hyeon:2017:TUR}
C. Hyeon and W. Hwang, 
Physical insight into the thermodynamic uncertainty relation using Brownian motion in tilted periodic potentials,
Phys. Rev. E {\bf 96}, 012156 (2017).

\bibitem{Horowitz:2017:TUR} 
J. M. Horowitz and T. R. Gingrich, 
Proof of the finite-time thermodynamic uncertainty relation for steady-state currents, 
Phys. Rev. E {\bf 96}, 020103(R) (2017).

\bibitem{Pigolotti:TURF}
S. Pigolotti, I. Neri, E. Roldán, and F. J\"ulicher, 
Generic Properties of Stochastic Entropy Production,
Phys. Rev. Lett. {\bf 119}, 140604 (2017).

\bibitem{Proesmans:2017:TUR}
K. Proesmans and C. V. den Broeck, 
Discrete-time thermodynamic uncertainty relation,
EPL {\bf 119}, 20001 (2017).


\bibitem{Hwang}
W. Hwang, and C. Hyeon,
Energetic costs, precision, and transport efficiency of molecular motors,
J. Phys. Chem. Lett. {\bf 9}, 513 (2018).
%


\bibitem{Bio}
R. Marsland III, W. Cui and J. M. Horowitz,
The thermodynamic uncertainty relation in biochemical oscillations,
J.  R.  Soc.  Interface {\bf 16}, (2019).  

\bibitem{Interacting}
S. Lee, C. Hyeon, and J. Jo, 
Thermodynamic uncertainty relation of interacting oscillators in synchrony,
Phys. Rev. E {\bf 98}, 032119 (2018).


\bibitem{Mayank}
M. Shreshtha and R. J. Harris,
Thermodynamic uncertainty for run-and-tumble-type processes,
EPL {\bf 126}, 40007  (2019). 

\bibitem{Garrahan:2017:TUR}
J. P. Garrahan, 
Simple bounds on fluctuations and uncertainty relations for first-passage times of counting observables,
Phys. Rev. E {\bf 95}, 032134 (2017).


\bibitem{Passage}
T. R. Gingrich and J. M. Horowitz,
Fundamental bounds on first passage time fluctuations for currents,
Phys. Rev. Lett. {\bf 119}, 170601 (2017).

\bibitem{Dechant:2018:TUR}
A. Dechant, 
Multidimensional thermodynamic uncertainty relations,
J. Phys. A: Math. Theor. {\bf 52}, 035001 (2019).


\bibitem{Pietzonka:2017:FiniteTUR} 
P. Pietzonka, F. Ritort, and U. Seifert,
Finite-time generalization of the thermodynamic uncertainty relation, 
Phys. Rev. E {\bf 96}, 012101 (2017).

\bibitem{Falasco}
G. Falasco, M. Esposito, and J.-C. Delvenne,
Unifying thermodynamic uncertainty relations,
arXiv:1906.11360.

\bibitem{SamuelssonM}
P. P. Potts and P. Samuelsson,
Thermodynamic uncertainty relations including measurement and feedback,
Phys. Rev. E {\bf 100}, 052137 (2019).

\bibitem{Koyuk:2018:PeriodicTUR}
T. Koyuk, U. Seifert, and P. Pietzonka, 
A generalization of the thermodynamic uncertainty relation to periodically driven systems,
J. Phys. A: Math. Theor. {\bf 52}, 02LT02 (2018).

\bibitem{Garrahan18}
K. Macieszczak, K. Brandner, and J. P. Garrahan, 
Unified thermodynamic uncertainty relations in linear response,
Phys. Rev. Lett. {\bf 121}, 130601 (2018).

\bibitem{Van}
T. Van Vu  and  Y. Hasegawa.
Uncertainty  relation  in  the  presence  of  information measurement and feedback control,
 arXiv:1904.04111


\bibitem{Hyst}
K. Proesmans and J. M. Horowitz,
Hysteretic thermodynamic uncertainty relation for systems with broken time-reversal symmetry,
J. Stat. Mech. 054005 (2019).

\bibitem{Gabri}
A. C. Barato, R. Chetrite, A. Faggionato, and D. Gabrielli, 
Bounds on current fluctuations in periodically driven systems,
New J. Phys. {\bf 20}, 103023 (2018).
 
\bibitem{Vu}
Y. Hasegawa and T. Van Vu,
Generalized thermodynamic uncertainty relation via the fluctuation theorem,
arXiv:1902.06376v3.

\bibitem{Hasegawa1} 
Y. Hasegawa and T. V. Vu, 
Uncertainty relations in stochastic processes: An information inequality approach, 
Phys. Rev. E {\bf 99}, 062126  (2019). 


\bibitem{Hasegawa2} 
M. L. Rosinberg and G. Tarjus, 
Comment on Thermodynamic uncertainty relation for time-delayed Langevin systems, 
arXiv:1810.12467.

\bibitem{Gingrich:2017}
T. R. Gingrich, G. M. Rotskoff and J. M Horowitz,
Inferring dissipation from current fluctuations, 
J.Phys. A: Math. Theor. {\bf 50} 184004 (2017). 

\bibitem{Sasa:TUR}
A. Dechant and S-i. Sasa, 
Current fluctuations and transport efficiency for general Langevin systems,
J. Stat. Mech. 063209 (2018). 

\bibitem{Saito}
K. Brandner, T. Hanazato, and K. Saito,
Thermodynamic bounds on precision in ballistic multiterminal transport,
Phys. Rev. Lett. {\bf 120}, 090601 (2018).

\bibitem{TUR-gupta}
D. Gupta and A. Maritan,
Thermodynamic uncertainty relations via second law of thermodynamics,
arXiv:1905.08854.

\bibitem{Udo:TURB}
H.-M. Chun, L. P. Fischer, and U. Seifert,
Effect of a magnetic field on the thermodynamic uncertainty relation, 
Phys. Rev. E {\bf 99}, 042128 (2019).

\bibitem{TURQ}
K. Ptaszynski,
Coherence-enhanced constancy of a quantum thermoelectric generator,
Phys. Rev. B {\bf 98}, 085425 (2018). 

\bibitem{BijayTUR}
B. K. Agarwalla and D. Segal,
Assessing the validity of the thermodynamic uncertainty relation in quantum systems,
Phys. Rev. B {\bf 98}, 155438 (2018).

\bibitem{TUR-bijay1}
S. Saryal, H. Friedman, D. Segal, and B. K. Agarwalla, 
Thermodynamic uncertainty relation in thermal transport, 
Phys. Rev. E {\bf 100}, 042101 (2019).


\bibitem{JunjieTUR}
J. Liu and D. Segal,
Thermodynamic uncertainty relation in quantum thermoelectric junctions,
Phys. Rev. E {\bf 99}, 062141 (2019).

\bibitem{SamuelssonQP}
S. Kheradsoud, N. Dashti, M. Misiorny, P. P. Potts, J. Splettstoesser, and P. Samuelsson,
Power, efficiency and fluctuations in a quantum point contact as steady-state thermoelectric heat engine,
Entropy {\bf 21}, 777 (2019).

\bibitem{Landi-PRL} 
A. M. Timpanaro, G. Guarnieri, J. Goold, and G. T. Landi, 
Thermodynamic uncertainty relations from exchange fluctuation theorems, 
Phys. Rev. Lett. {\bf 123}, 090604 (2019).

\bibitem{VanTUR}
Y. Hasegawa and T. Van Vu,
Fluctuation theorem uncertainty relation,
Phys. Rev. Lett. {\bf 123}, 110602 (2019). 

\bibitem{Goold}
G. Guarnieri, G. T. Landi, S. R. Clark, and J. Goold,
Thermodynamics of precision in quantum non-equilibrium steady states,
arXiv:1901.10428.

\bibitem{Horowitz:2019:TUR}
J. M. Horowitz and T. R. Gingrich
Thermodynamic uncertainty relations constrain non-equilibrium fluctuations,
Nat. Phys. (2019).

\bibitem{fluct1}
M. Esposito, U. Harbola, and S. Mukamel,
Nonequilibrium fluctuations, fluctuation theorems, and counting statistics in quantum systems,
Rev. Mod. Phys. {\bf 81}, 1665 (2009).

\bibitem{fluct2}
M. Campisi, P. H\"anggi, and  P. Talkner,
Colloquium: Quantum fluctuation relations: Foundations and applications,
Rev. Mod. Phys. {\bf 83}, 771 (2011).

\bibitem{campisi-measurement}
M. Campisi, P. Talkner, and P. H\"anggi,
Influence of measurements on the statistics of work performed on a quantum system,
Phys. Rev. E {\bf 83}, 041114 (2011).

\bibitem{JarzW}
C. Jarzynski and D. K. Wójcik,
Classical and quantum fluctuation theorems for heat exchange,
Phys. Rev. Lett. {\bf 92}, 230602 (2004).


\bibitem{Wei} B. B. Wei,
Relations between heat exchange and R\'enyi divergences,
Phys. Rev. E {\bf 97}, 042107 (2018).


\bibitem{supp}
See supplemental Material File


\bibitem{ancilla-1}
L. Mazzola, G. De Chiara, and M. Paternostro,
Measuring the characteristic function of the work distribution,
Phys. Rev. Lett. {\bf 110}, 230602 (2013).

\bibitem{ancilla-2}
R. Dorner, S. R. Clark, L. Heaney, R. Fazio, J. Goold, and V. Vedral,
Extracting quantum work statistics and fluctuation theorems by single-qubit interferometry,
Phys. Rev. Lett. {\bf 110}, 230601  (2013).


\bibitem{ancilla-3}
M. Campisi, R. Blattmann, S. Kohler, D. Zueco, and P. Hanggi,
Employing circuit QED to measure non-equilibrium work fluctuations,
New J. Phys. {\bf 15}, 105028 (2013).



\bibitem{Bijay-expt} 
S. Pal, T. S. Mahesh, and B. K. Agarwalla, 
Experimental demonstration of the validity of the quantum heat-exchange fluctuation relation in an NMR setup,
 Phys. Rev. A {\bf 100}, 042119 (2019).

\bibitem{XFT-theory}
G. T. Landi and D. Karevski,
Fluctuations of the heat exchanged between two quantum spin chains,
Phys. Rev. E {\bf 93}, 032122 (2016).

\bibitem{XFT-agarwalla}
B. K. Agarwalla, H. Li, B. Li, and J.-S. Wang,
Exchange fluctuation theorem for heat transport between multiterminal harmonic systems,
Phys. Rev. E {\bf 89}, 052101 (2014).

\bibitem{Bijay12}
B. K. Agarwalla, B. Li, and  J.-S. Wang,
Full-counting statistics of heat transport in harmonic junctions: Transient, steady states, and fluctuation theorems,
Phys. Rev. E {\bf 85}, 051142 (2012).

\bibitem{Saito07}
K. Saito and A. Dhar,
Fluctuation theorem in quantum heat conduction,
Phys. Rev. Lett. \textbf{99}, 180601 (2007).

\bibitem{Lutz_2018}
T. Denzler and E. Lutz,
Heat distribution of a quantum harmonic oscillator,
Phys. Rev. E {\bf 98}, 052106  (2018).

\bibitem{SaitoUts}
K. Saito and Y. Utsumi,
Symmetry in full counting statistics, fluctuation theorem, and relations among nonlinear transport coefficients in the presence of a magnetic field,
Phys. Rev. B \textbf{78},  115429 (2008).


\bibitem{comment}
In the weak coupling limit the second cumulant (\ref{eq:analM}) can be organized as (using $x \coth(x) \geq 1$)
\bea
\langle Q^2 \rangle^{c}_{\tau} &=& (h \nu_0)^2 {\cal T}_{\tau}(J) \Big( f_1 (1\!-\!f_2) \!+\! f_2 (1\!-\!f_1)\Big) \nonumber \\
&=&  (h \nu_0)^2  \coth\Big(\frac{\Delta \beta \, h \nu_0}{2}\Big) {\cal T}_{\tau}(J) (f_2 - f_1) \nonumber \\
&\geq & \frac{2}{\Delta \beta} h \nu_0 {\cal T}_{\tau}(J) (f_2 - f_1) = \frac{2}{\Delta \beta} \langle Q \rangle_{\tau},
\eea
%
proving that the S-TUR is satisfied even far from equilibrium.







\end{thebibliography}
\end{document}